# Strong Influence of Phonons on the Electron Dynamics of $Bi_2Sr_2CaCu_2O_{8+\delta}$

G.-H. Gweon[*], S. Y. Zhou[†], and A. Lanzara[†§]

*\* Lawrence Berkeley National Laboratory, Berkeley, CA 94720, U.S.A.*
*† Department of Physics, University of California, Berkeley, CA 94720, U.S.A.*
*§ Materials Sciences Division, Lawrence Berkeley National Laboratory, Berkeley, CA 94720, U.S.A.*

The sudden change of the velocity, so-called "kink", of the dispersing peak in angle resolved photoelectron spectroscopy is a well-known feature in the high temperature superconducting cuprates. Currently, the origin of the kink is being much debated, but a consensus has not emerged yet. Here, we present a study of the momentum evolution of the kink structure from the nodal region towards the anti-nodal region, for optimally doped $Bi_2Sr_2CaCu_2O_{8+\delta}$ sample. We show that the observed temperature dependence of the kink structure in both regions of the momentum space is consistent with a scenario in which phonons contribute strongly to the kink.

In addressing central questions about the electronic structure in the high temperature superconducting cuprates, angle resolved photoelectron spectroscopy (ARPES) has played a major role. Its key contributions include Fermi surface geometry, superconducting gap symmetry, and pseudo-gap in the normal state [1,2]. Currently, one of the most controversial issues in ARPES on the cuprates is the question surrounding the "kink", an abrupt change in the group velocity ($\partial E/\partial k$) of the ARPES peak [3-5]. The importance of the kink lies in the fact that it represents the lowest energy scale, except for the superconducting gap itself, known in the electron dynamics of the cuprates [1,2].

The kink may be generally interpreted as the presence of an energy scale [6]. In particular, the kink has been often explained in terms of coupling of the electron to a bosonic collective mode [7], and in the case of simple metals the mode has been easily identified with phonons [8]. In the case of the cuprate superconductors, a consensus on the origin of the kink structure has not emerged yet. Two main lines of thoughts on its origin have been proposed and are currently under debate: coupling to a magnetic mode [9-11] and coupling to phonons [4,5]. The first scenario invokes coupling to the well-known magnetic mode, observed in the inelastic neutron scattering experiment [12], and often referred as the "INS mode". Because (for the p-type cuprates) the wave vector of the INS mode, $(\pi,\pi)$, connects Fermi surface points near one anti-nodal point, e.g. $(-\pi,0)$, to those near another, $(0,\pi)$, the contribution of the INS mode is expected to be most relevant near the anti-nodal point of the Brillouin zone. The contribution, if any, near the nodal direction, i.e. $(0,0) \to (\pi,\pi)$ direction, is unclear. While all ARPES groups in favor of this scenario agree that the INS mode is important near the anti-nodal point, there is no good agreement regarding its importance near the nodal direction. Opinions range from the INS mode being responsible for the kink near the anti-nodal point $(\pi, 0)$ only [11] to the INS mode being responsible also for the nodal kink [10]. In arguments advancing the INS mode scenario, the temperature dependence of the kink is used as an important element. Here, the kink above T* (pseudo-gap temperature) and the kink below T* are distinguished: the former is interpreted in the framework of the marginal Fermi liquid (MFL) and the latter attributed to the INS mode, known to exist only below T*. The second view associates the nodal kink to a strong electron-phonon interaction [4,5]. This view is borne out of the observation that the kink is observed universally, i.e. at similar energy among all different materials measured by ARPES so far, regardless of the temperature and the presence of the INS mode.

The purpose of this paper is to look at this issue more closely, in order to assess the relative strength of the arguments used for each of these scenarios. For this, we studied nodal cuts and off-nodal cuts, up to $\approx$ half way towards the $(\pi,0)$ point. For all these cuts, we show that the normal state kink can *not* be interpreted within the MFL framework. It follows that a unified description of the normal state kink and the superconducting state kink is most natural, and that the phonon dynamics is a significant part (if not all) of the kink dynamics, for nodal *and* anti-nodal regions.

The experiments were carried out at beam line 10 of the Advance Light Source of the Lawrence Berkeley National Laboratory. Photon energy used was 33 eV. Data were taken with energy resolution of 15 meV FWHM and angular resolution of 0.15 degrees. Sample is an optimally doped $Bi_2Sr_2CaCu_2O_{8+\delta}$ with $T_c$ = 92 K. Measurement temperatures were 25 K and 100 K. At 100 K, the pseudo-gap is absent in our data, i.e. $T^* < 100$ K. This implies $T^* \approx T_c$, reasonable for an optimally doped sample. Therefore, in this paper, we will use expressions "above $T_c$" and "above $T^*$" interchangeably, both meaning "at 100 K". As a matter of convention, we will use the expression "high energy" to mean "high binding energy" and "low energy" to mean "low binding energy" for photo-electrons. All of our cuts are parallel to the nodal direction, and they will be numbered using the angle offset from the nodal direction. For example, cut 0 will mean a nodal cut, and cut 6 will mean a cut 6 degrees displaced from the nodal cut.

Fig. 1 shows the ARPES data taken along the nodal direction (cut 0) and $\approx$ half way between the nodal direction and the $(\pi,0)$ point (cut 6). The precise locations of the cuts are given in the inset. Panels a and b show gray scale maps of the raw spectral intensity in the energy-momentum space (E, **k**). Both low temperature (25 K) and high temperature (100 K) data are shown. In panel c, we report the momentum distribution curve (MDC) dispersions extracted from the MDC peak positions, using Lorentzian line shape fitting.

Our data highlight the following points. First, the kink is observed for *all* dispersions, i.e. there is a crossover from a high velocity for E < -100 meV to a low velocity for E > -50 meV (the velocity is proportional to $\partial E/\partial k$, and simply corresponds to the slope of the dispersion curve). The dashed lines, guides to the eye for the high energy dispersion at high temperature, are included to convey this point clearly. Second, there is much greater temperature dependence in the off-nodal cut than in the nodal cut. For the nodal cut (solid lines), the dispersion is nearly temperature independent, while for the anti-nodal cut (dashed lines) the dispersion at low temperature is quite different from that at high temperature. Third, the low temperature data for the off-nodal cut shows a superconducting gap opening. The back bending of the MDC dispersion near $E_F$ in the 25 K curve of Fig. 1c is a well-known effect of the gap opening in the MDC dispersion. Our data are in good agreement with those reported for similar cuts in the literature [9-11].

The central issue regarding the controversies surrounding the kink can be summarized as "*is the nature of the kink the same above and below $T_c$?*". Here, we can already make simple intuitive observations. First, for the nodal cut, the noted similarity between the high temperature and the low temperature kink structures, argue strongly for the single origin of the two. In detail, there is a noticeable softening of the kink at high temperature. Within the commonly used Eliashberg-type electron-boson coupled model, this softening occurs due to the finite temperature quantum statistics and it was shown that a significant softening occurs already at T ≈ 0.1 Ω where Ω is the energy of the boson mode [13]. A temperature of 100 K and Ω of 70 meV, i.e. the approximate kink position extracted from our data, satisfy this condition. Therefore, it seems quite reasonable that a single origin explains the kink in the nodal direction. Second, the case of the off-nodal kink is clearly more complex. Namely, the kink below $T_c$ is much pronounced than the kink above $T_c$.

To gain more insight into the kink dynamics, we show, in Fig. 2a, the MDC width as a function of energy, obtained within the same fitting procedure used to extract the dispersions in Fig. 1. It is evident from the data that the MDC width shows a sudden drop between 50-80 meV in the superconducting phase. More importantly, a similar, but weakened, drop is observed in the normal phase for *all* cuts (e.g. see dashed lines for cut 0 and cut 6). Incidentally, the small peak observed in the low temperature data (binding energy 30 meV or less) is another signature of the superconducting gap opening.

The data of Fig. 2a strongly challenges the view that the normal state is described as a MFL. In this view, the MFL theory gives the imaginary part of self energy ImΣ ∝ ω [14], which is taken to be proportional to the MDC width, within the **k**-independent self energy approximation. However, our high temperature data show that this is not the case, since a step is observed. Such a step invalidates the very assumption that goes into the MFL theory, namely the absence of an energy scale. No signature of pseudogap is present in the (π,0) spectrum at high temperature, dispelling potential doubt that our data are taken below the T*, where the INS peak develops [12]. Careful re-examination of the published data, for which the MFL self energy claim was made [9,10], leads us to conclude that the better statistical quality of our data is the key to our finding. The current data fall in the same range with the other data, but we can now better resolve the small drop at 50-80meV.

Therefore, our results point to a scenario that *a similar energy scale exists both below and above $T_c$* and the strength of the kink simply increases as the temperature decreases.

For a more detailed analysis of the kink energy position, we study ReΣ(ω), shown in Fig. 2b. The kink energy is of fundamental importance in identifying the nature of the bosonic excitation coupled to the electron. For example, in the Eliashberg-type model, the characteristic energy of the boson is the kink energy at which the ReΣ(ω) has a peak and the ImΣ(ω) has a drop (by the Kramers-Kronig relation). It must be noted that, indeed, the ReΣ(ω) analysis of Fig. 2b shows a peak at a similar energy where the MDC width, associated with ImΣ(ω), shows a drop. This justifies the use of self energy for the kink analysis. We choose ReΣ(ω) for a numerical analysis of the kink position since the data for ReΣ(ω) are technically better suited for this purpose than the data for ImΣ(ω). However, the results of the ReΣ(ω) analysis must be interpreted carefully without overlooking the weakness of two key underlying assumptions. The first assumption is the **k**-independent Σ. The Lorentzian-like line shape of *each* MDC gives some justification for this assumption in a narrow range of **k** corresponding to the MDC width, but this is *by no means* a guarantee that Σ is **k**-independent in the entire **k** range. In fact, we see below that ReΣ(ω)'s vary significantly for different cuts. Second, a form of the non-interacting band dispersion ε(**k**) *must be assumed*. For this, we use a line connecting the two points (at -200 meV and $E_F$) of the MDC dispersion at high temperature. Since ReΣ(ω) is generally a small fraction of the dispersion ε(**k**) in magnitude, the inherent uncertainty of ε(**k**) leads to a large uncertainty in ReΣ(ω). For example, we find that the kink energy increases by ≈ 10 meV if we use a tight binding form of ε(**k**), rather than a linear ε(**k**), in the current example. This is because the tight binding band has a roughly parabolic dispersion, giving a bigger ReΣ(ω) for large values of ω than a linear ε(**k**) would. In this context, the strong argument for the INS mode scenario made using the temperature dependence of the absolute value of ReΣ(ω) [10] loses much of its strength.

The following observations can be made on ReΣ at low temperature shown in Fig. 2b. First, the data show a peak, located between 60-80 meV. The position of the peak shifts by ≈10 meV going from cut 0 to cut 6 (see arrows). Second, in addition to the peak, ReΣ contains significant strength in the high energy part. This is especially true for off-nodal cuts (e.g. cut 6). This implies that the broad peak at this high energy does *not* represent an incoherently propagating quasi-free electron but an electron still *heavily dressed by interaction*, including large electron-phonon interaction as recently suggested [15]. Incidentally, the negative ReΣ near $E_F$ comes from the bending back noted in Fig. 1, and therefore arises due to the superconducting gap opening.

Now, we discuss the temperature dependence of ReΣ. At low temperature, ReΣ is enhanced greatly for cut 6, which is the angle at which the superconducting gap is the largest. The large low T enhancement of ReΣ is also consistent with the large enhancement of the MDC width (Fig. 2a) and the kink structure (Fig. 1). In terms of line shape, ReΣ at high temperature shows a more rounded peak, compared to the low temperature data. Note that this rounded peak has been often fit with the MFL theory, but this cannot be justified in a more complete picture that we present here. While ReΣ alone does not necessarily rule out the MFL theory (with a freely adjustable ultraviolet energy cutoff parameter), the clear drop in the MDC width does, as we explained earlier in this paper.

In Fig. 3, we summarize the kink position obtained from ReΣ using a quadratic curve fit near the maximum of the peak (circles) and from the dispersion using a two-line intersection method (diamonds). We will discuss the results obtained by the first method mainly. Values obtained by the second method are discussed near the end of the paper. Fig. 3a shows the kink position as a function of angle at low temperature. As discussed above, the kink position gradually decreases by ≈10 meV (from ≈75 meV to ≈65 meV) as **k** moves away from the nodal direction. Fig. 3b shows the kink positions for high temperature. Note larger error bars in this case, due to the broadness of the high temperature humps in ReΣ. Despite the large error bar, the kink energy is apparently larger at high temperature. Within the Eliashberg-type model, temperature broadening is known to give rise to this apparent shift of the kink energy [13]. However, more important effect appears to be that the peak at low energy broadens at high temperature and merges with the "tail" at high energy to cause this apparent kink energy shift.

To summarize, we have shown ARPES data below and above $T_c$ for an optimally doped sample of $Bi_2Sr_2CaCu_2O_{8+\delta}$ along several cuts covering nodal region and near-anti-nodal region. The key findings are as follows. (1) A drop in the MDC width is observed both below and above $T_c$. (2) ReΣ shows a peak near the energy position of the

MDC width drop both below and above $T_c$. (3) The kink structure in the nodal direction is only slightly temperature dependent, involving a softening of the kink at high temperature. (4) The kink energy is momentum dependent and decreases by ≈10 meV going from the nodal direction (cut 0) toward the anti-nodal point (cut 6). (5) The temperature induced change for off-nodal cut, e.g. cut 6, is much more dramatic than the nodal cut. At low temperature, the peak (step) becomes more distinct in ReΣ (MDC width). Both the MDC width and ReΣ also increase greatly in magnitude at low temperature.

The findings (1), (2) and (3) are strong indications that the kinks above and below $T_c$ are of the same origin. The presence of a kink above $T_c$ clearly suggests that the involved dynamics is not that of the INS mode. In view of previous results, especially that of the strong isotope effect [15], it appears quite plausible that the kink dynamics is significantly phonon-related. On the other hand, the findings (4), (5) appear to indicate that the kink near the (π,0) point, i.e. the kink for cut 6, may require a somewhat different involvement of phonons at low temperature.

Here we propose a new scenario, where both the nodal and antinodal kinks can be explained in terms of coupling of quasiparticles to the lattice. In this scenario, the high temperature data show the kink dynamics involving more than one characteristic phonon energy, where the kink energy is an energy cutoff and corresponds to the highest phonon frequency involved, which in this case corresponds to the longitudinal half-breathing phonon of in plane oxygen [16-18]. At low temperatures, in the momentum space region where the superconducting gap opens up, coupling to lower energy phonons is enhanced, i.e the kink position scales to lower frequency, as seen for the off nodal cuts (figure 1). The enhancement occurs because the opening of the superconducting gap induces an increase of the density of states at the gap edge, and thus increases effective coupling to phonons whose frequency matches the excitation across the gap. Another way to look at it is to think in terms of a resonance mechanism involving the proximity of the phonon energy scale and the particle-hole excitation across the superconducting gap (maximum energy 2Δ) [19-21].

A comparison of the kink energy position with the Raman and neutron scattering data leads us to conclude that, for antinodal quasiparticles, coupling to lower energy phonon is enhanced below $T_c$. We identify the low energy phonons as the tilting phonons (the lowest oxygen phonon frequency), also known as the buckling phonon [22]. The tilting angle is defined as the angle by which the in plane oxygen atoms are "forced" out of the plane of the copper atoms [22]. It is important to point out that the tilting phonon and the half breathing phonon are related to each other [18] and in particular the tilting phonon is the main phonon involved in the low temperature structural transition of certain cuprates [18, 23]. These transitions are associated with the strain on the $CuO_2$ plane and the tilting of the plane is a way for the system to release such strain [23-25].

This proposed scenario can explain the recently observed trend of the kink strength as a function of the number of the $CuO_2$ layer [11]. First, the almost temperature independent off-nodal kink in the single layer Bi-compound is the result of too small Δ (≈10 meV) to induce resonance. Second, the weakening of the normal state kink as the number of $CuO_2$ planes per unit cell increases can be understood as the lessening of the strain per $CuO_2$ plane. To take this scenario further, one might speculate on the link between the tilting phonon and the nested Fermi surface region near the anti-nodal point, as well as a possible link between this anisotropic phonon coupling to the anisotropic superconducting gap [26, 27].

For a more quantitative analysis in the future, it is important to point out that the absolute value of the kink position may be subject to a rather large uncertainty, although the difference in energies of the nodal and the anti-nodal kinks at low temperature is robust ($\approx$10 meV). At low temperature, the kink energy values, as determined from the Re$\Sigma$ peak position, are: $\approx$75 meV for cut 1 and $\approx$65 meV for cut 6. In addition to the numerical uncertainty ($\approx\pm$5 meV), these values are subject to an additional uncertainty ($\approx\pm$10 meV) due to the uncertainty of the bare band dispersion. Furthermore, if we adopt a different method to extract the kink energy position, somewhat different answers can be obtained. For example, the two straight-line method [4] (diamonds in Fig. 3) applied to the MDC dispersion curves gives the nodal kink energy of $\approx$60 meV and the off nodal (cut 6) energy of $\approx$50 meV at low temperature, and even smaller value ($\approx$40 meV) for the off-nodal kink at high temperature. Overall, these uncertainties imply that a more detailed study is needed to carefully characterize the normal state kink.

The importance of the phonon in the electron dynamics pointed out here is in agreement with previous studies [4,5], except, however, that the phonons involved in the off-nodal kink structure were not identified before. We hope that our findings lead the community to reach closer to the solution of the high temperature superconductivity, which clearly requires a careful modeling including not only electron-electron interaction but also electron-phonon interaction.

**Acknowledgement**

We would like to acknowledge useful and insightful discussion with N.L. Saini and D. H. Lee, A. K. Müller, Z. Hussain, T. Sasagawa and H. Takagi. This work was supported by the Director, Office of Science, Office of Basic Energy Sciences, Division of Materials Sciences and Engineering, of the U.S. Department of Energy under Contract No. DE-AC03-76SF00098.

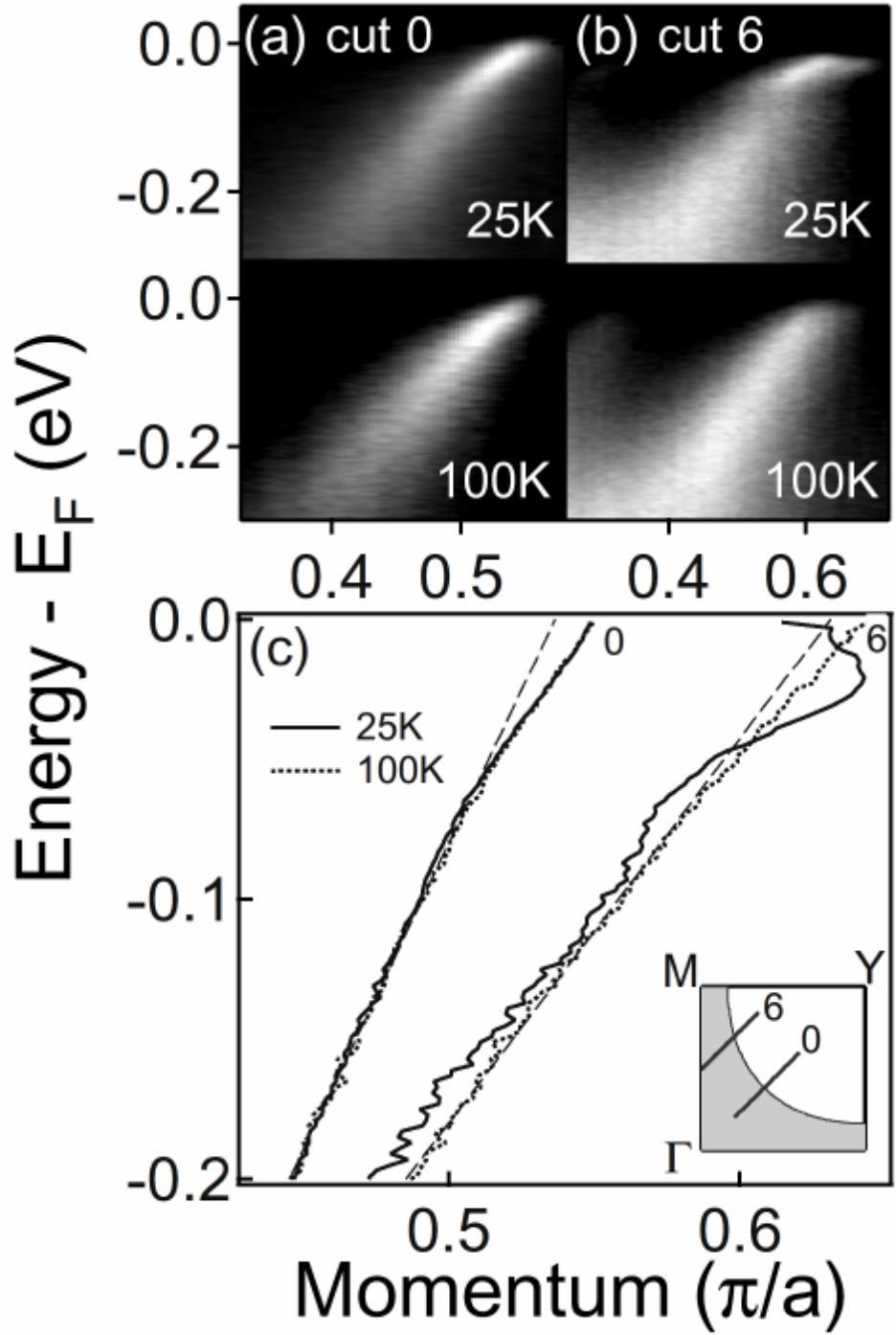

**Fig. 1** (a) ARPES intensity map along cut 0 and (b) cut 6 on optimally doped $Bi_2Sr_2CaCu_2O_{8+\delta}$. The cut locations are given in inset of panel c. (c) MDC dispersions obtained by MDC fits of the data in panels a and b with Lorentzian curves. Dashed lines are guides to the eye for the high energy dispersions. Momentum axis is defined by momentum value projected to the ΓY axis.

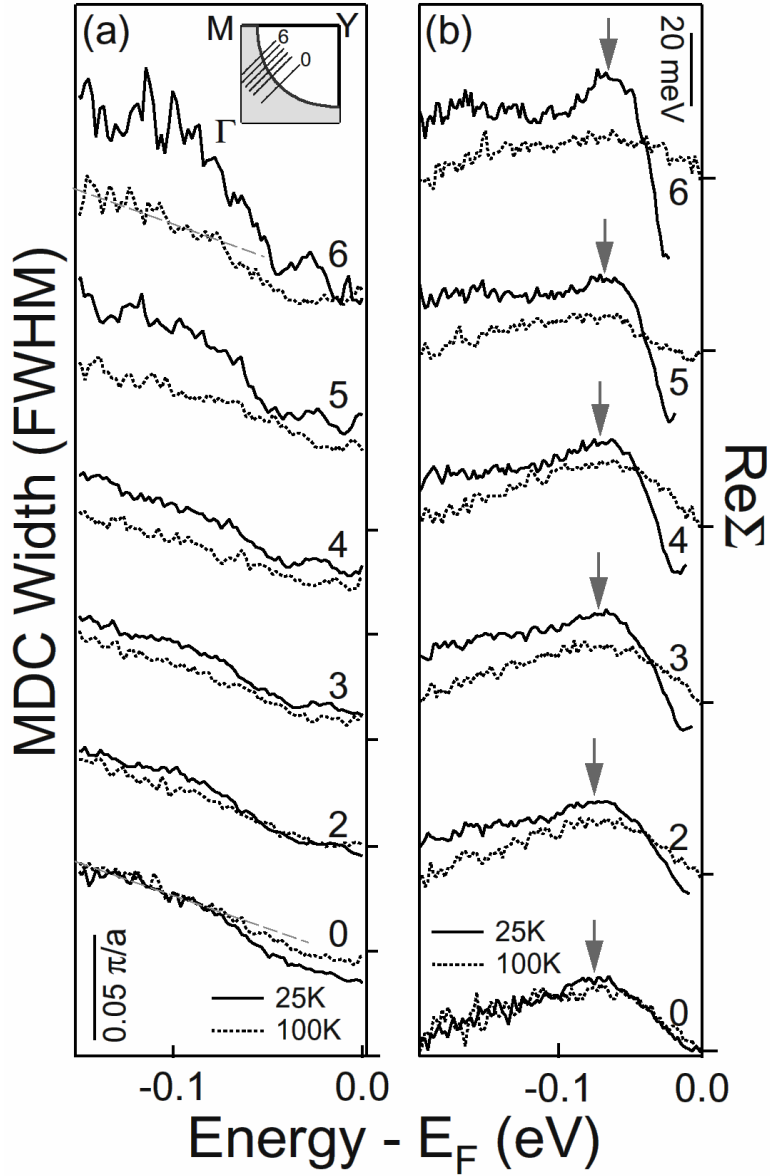

**Fig. 2** (a) MDC width extracted from the MDC fit of the data for cuts 0 to 6. Dashed lines are given as guides to the eye for the high energy part of the curves for cut 0 and 6. A drop in the MDC width is clearly observable, for all cuts at both temperatures. (b) ReΣ(ω) extracted from the data, as described in the text. Arrows mark the peak positions of low temperature curves obtained by parabolic fits. For both panels, data are shifted vertically for ease of view, and the six small tick marks on the right side of the panel show the position of the zero of the vertical axis for each of the six cuts shown, from bottom (cut 0) to top (cut 6).

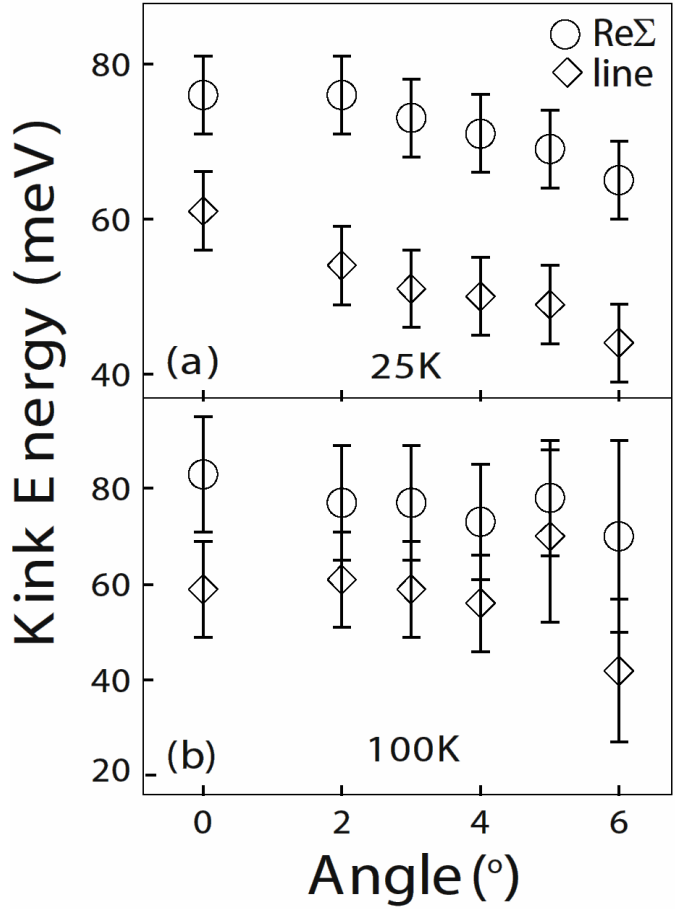

**Fig. 3** Kink energies extracted from the data for (a) low temperature and (b) high temperature. Two different methods are used, one using the peak position of ReΣ(ω) (circles) and the other using a two-line intersection method on the MDC dispersion curves (diamonds) [4].